\documentclass[%
 reprint,
 amsmath,amssymb,
 aps,
 pra,
 showkeys,
floatfix,
]{revtex4-2}
\usepackage[utf8]{inputenc}
\usepackage{graphicx}
\usepackage{dcolumn}
\usepackage{bm}
\usepackage{bbold}
\usepackage{physics}
\usepackage[colorlinks=true, allcolors=blue]{hyperref}
\usepackage{siunitx}
\usepackage{mathdots}
\usepackage[dvipsnames]{xcolor}
\usepackage{comment}
\DeclareSIUnit\dBm{dBm}

\newcommand{\matr}[1]{\bm{#1}}
\newcommand{\matrx}[1]{\bm{\mathcal{#1}}}
\newcommand{\identity}{\mathbb{1}}

\definecolor{drkgreen}{rgb}{0.0, 0.5, 0.0}
\definecolor{violet}{rgb}{0.5, 0.0, 1.0}

\begin{document}
\preprint{APS/123-QED}
\title{Solving the inverse parametric problem}%
\author{Michele Cortinovis}%
\author{Fabio Lingua}\email{lingua@kth.se}%
\author{David B. Haviland}%
\affiliation{Department of Applied Physics, KTH Royal Institute of Technology, SE-10691 Stockholm, Sweden}%
\date{\today}

\begin{abstract}

We present a method to calculate the frequency components of a pump waveform driving a parametric oscillator, which realizes a desired frequency mixing or scattering between frequency modes. The method is validated by numerical analysis and we study its sensitivity to added Gaussian noise. A series of experiments apply the method and demonstrate its ability to realize complex scattering processes involving many modes at microwave frequencies, including non-reciprocal mode circulation. We also present an approximate method to dynamically control mode scattering, capable of rapidly routing signals between modes in a prescribed manner. These methods are useful tools for encoding and manipulating continuous variable quantum information with multi-modal Gaussian states.
\end{abstract}


\maketitle



\section{\label{sec:introduction}Introduction}

Parametric oscillators are a universal paradigm for a variety of applications across physics.  
They are extensively used in quantum technologies for quantum limited amplification~\cite{caves_quantum_2012}, coupling qubits~\cite{bertet_parametric_2006}, quadrature squeezing~\cite{yurke_squeezed-state_1987, yurke_lownoise_1996, castellanos-beltran_widely_2007, castellanos-beltran_amplification_2008, mallet_quantum_2011, malnou_optimal_2018, aumentado_superconducting_2020,liao_parametric_2011}, as well as engineering entanglement through frequency mixing and second-harmonic generation, at microwave~\cite{pfister_multipartite_2004, eichler_observation_2011, andersson_squeezing_2022, esposito_observation_2022, jolin_multipartite_2023, petrovnin_generation_2023, lingua_continuous-variable_2025}, and at optical wavelengths~\cite{cerullo_ultrafast_2003,yokoyama_ultra-large-scale_2013, chen_experimental_2014}.

\begin{figure}
    \centering
    \includegraphics[width=\columnwidth]{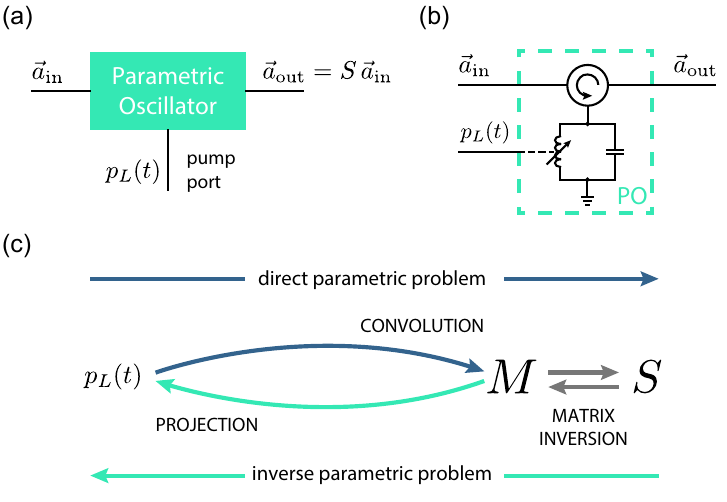}
    \caption{(a) General I/O relation of a parametric oscillator where the output is determined by the scattering matrix $\matr{S}$, controlled by the pump $p_L(t)$. 
    (b) Example of a parametric oscillator, the pump port modulates the inductance of an LC oscillator. (c) We define direct problem when the knowledge of $p_L(T)$ allows to compute the scattering matrix $\matr{S}$, inverse problem when from a target $\matr{S}$ one recover the pump signal that realizes it. }
    \label{fig:PO-setup}
\end{figure}

A parameter of an oscillator is externally controlled by a coherent drive known as the \emph{pump}.  
The pump can be a continuous wave consisting of a single frequency, or a periodic waveform consisting of a sum of multiple tones.
The frequency, amplitude, and phase of these tones define the pump waveform, effectively controlling the parametric process.
Given a particular pump waveform $p_L(t)$, a direct computation of the equation-of-motion matrix $\matr{M}$ yields the scattering matrix $\matr{S}$ which describes how an input $a_\text{in}(t)$ is transformed to an output $a_\text{out}(t)$ (see fig. \ref{fig:PO-setup}a)~\cite{gardiner_input_1985, naaman_synthesis_2022}. 
We call this computation the \emph{direct parametric problem} (see fig. \ref{fig:PO-setup}c), useful for describing parametric amplification~\cite{mollow_quantum_1967}, entanglement generation~\cite{pfister_multipartite_2004, eichler_observation_2011, andersson_squeezing_2022, esposito_observation_2022, jolin_multipartite_2023}, multi-mode squeezing and cluster state generation~\cite{yokoyama_ultra-large-scale_2013, chen_experimental_2014, petrovnin_generation_2023,lingua_continuous-variable_2025}, signal routing and frequency conversion~\cite{ranzani_graph-based_2015, naaman_synthesis_2022, hernandez_control_2024}, even non-reciprocal~\cite{lecocq_nonreciprocal_2017, bock_nonreciprocal_2025}.

More interesting in many applications is a solution to  the \emph{inverse parametric problem}: given a desired scattering matrix, what is the required pump waveform?
Previous attempts to solve this problem have relied on optimization methods such as automated pump shaping~\cite{arzani_versatile_2018}, discovery of coupled-mode networks~\cite{landgraf_automated_2025}, or machine learning strategies~\cite{namdar_spectro-temporally_2025}. 
To our knowledge, a general and systematic framework for solving the inverse parametric problem has remained an open challenge.

Here we propose an exact solution to this problem in the frequency domain.  
Our method relies on mode-coupling theory~\cite{naaman_synthesis_2022} which is traditionally used in network synthesis, where coupled resonators realize scattering between physically distinct ports.
We use the same theoretical framework to describe scattering between the frequency modes of a transmission line reflected from a single port of a multi-frequency pumped parametric oscillator. 
The method handles systems with high pump degeneracy, where pump amplitudes do not map directly onto individual Hamiltonian terms.
In this picture each pump frequency is an element of an orthogonal basis spanning the Hilbert space of all possible matrices $\matr{M}$. 
Defining a suitably normalized inner product on this Hilbert space, we directly obtain the amplitude and phase at each pump frequency  as a projection of the $\matr{M}$ that gives the desired output.
We implement the method numerically and assess  its robustness against noise, demonstrating the ability to reconstruct randomly generated scattering matrices. 
We use the method to design circulation between multiple frequency modes described by a given ideal target scattering matrix. 
We also demonstrate another method which encodes arbitrary information in the output modes by dynamically changing the pump waveform.

\section{Parametric Oscillator and Inverse Problem} \label{sec:theory}


A general parametric oscillator is described by the time-dependent Hamiltonian
\begin{equation}\label{eq:HPO}
    \hat{\mathcal{H}}_\text{PO} =
    \frac{\omega_0}{2}\left(A^\dag A + A A^\dag\right)
    + \frac{\omega_0}{2}\, p_L(t)\left(A + A^\dag\right)^2,
\end{equation}
where $A$ and $A^\dag$ are the annihilation and creation operators of the oscillator mode with resonant frequency $\omega_0$, which is modulated by a pump $p_L(t)$.
We consider pump waveforms that are periodic over a time interval $T$, expressed as a coherent superposition of tones with  frequencies $\Omega_k$, amplitudes $p_k$, and phases $\phi_k$:
\begin{equation}\label{eq:pump}
    p_L(t) = \sum_k p_k \cos(\Omega_k t + \phi_k)
           = \sum_k \left(g_k e^{-i\Omega_k t} + g_k^* e^{i\Omega_k t}\right),
\end{equation}
where $g_k = \tfrac{p_k}{2} e^{i\phi_k}$ are the complex pump amplitudes.

The parametric oscillator is coupled to an external bath of modes, allowing an input–output description of its dynamics. 
In superconducting implementations, this coupling typically occurs through input and output transmission lines, as illustrated in Fig.~\ref{fig:PO-setup}(b). 
The oscillator mode $A$ ($A^\dag$) is coupled to the continuum of propagating modes in the transmission lines, described by the field operators $a(\omega)$ and $a^\dag(\omega)$. 
The measurement time window $T$ sets the frequency resolution $\Delta = 1/T$, so that the field in the transmission line is represented by a discrete set of orthogonal frequency modes $a_m$ and $a_m^\dag$.
A convenient way to model the dynamics of the parametric oscillator is to expand Eq.~\eqref{eq:HPO} in the frequency-mode basis of the transmission lines, $A^{(\dag)} = \sum_m a_m^{(\dag)} e^{-i\omega_m t}$ (see Ref.~\cite{hernandez_control_2024} for a detailed derivation). 

This expansion transforms the time-dependent Hamiltonian into a time-independent system of $N$ coupled harmonic oscillators~\cite{ranzani_graph-based_2015,naaman_synthesis_2022,hernandez_control_2024}. 
The resulting equation of motion (EOM) for the $m^\text{th}$ mode in the frequency domain reads
\begin{equation}\label{eq:EOM_f}
    i(\omega_m + \tilde{\omega}_0)\, a_m
    + i\!\sum_n\!\left(g_{mn} a_n + g_{mn}^* a_n^\dag\right)
    = \sqrt{\gamma}\, a_{\text{in},m},
\end{equation}
where $\tilde{\omega}_0 = \omega_0 - i\gamma/2$ and $\gamma$ denotes the coupling rate to the transmission line, assumed constant for all modes. 

The parametric pump enables intermodulation, or frequency mixing between modes through the couplings $g_{mn} = \sum_k g_k\, \delta_{m,n\pm k}$, 
which connect modes satisfying the condition $\omega_m \pm\omega_n= \pm \Omega_k$. 
This mixing is a direct consequence of multiplication with $p_L(t)$ in Eq.~\ref{eq:HPO}, which is a convolution in the frequency domain (see~\cite{hernandez_control_2024} for further details).
The validity of this framework is contingent upon the assumption that (i) linear regime holds, $\mathcal{H}_{\text{PO}}$ is quadratic and intrinsic higher order (i. e. Kerr) terms are considered negligible. 
Any presence of quartic or higher order terms in $\mathcal{H}_{\text{PO}}$ would create intermodulation processes through four-wave or higher mixing that cannot be described by~\eqref{eq:EOM_f} and the method that follows hereafter.
(ii) the PO is operated in the stiff pump regime, i.e. the pump is sufficiently strong compared to the input-output signals so that pump-depletion can be neglected. 
(iii) The frequency modes $\omega_m$ lie well within the bandwidth of the PO.

The equations of motion can be compactly written in matrix form as
\begin{equation}\label{eq:EOM_matrix}
    -i\gamma\, \matr{M}\, \vec{a} = \sqrt{\gamma}\, \vec{a}_\text{in},
\end{equation}
where $\vec{a} = (a_1, a_1^\dag, \dots, a_N, a_N^\dag)^{T}$ collects all mode operators satisfying the bosonic commutation relations $[a_m, a_n^\dag] = \delta_{mn}$. 
In the classical limit $a_m^{(\dag)} \!\to\! a_m^{(*)} \!\in\! \mathbb{C}$. 
Following Ref.~\cite{ranzani_graph-based_2015}, we normalize $\matr{M}$ by $\gamma$.

We distinguish two types of pumps, low frequency (LF, sometimes called beam-splitter or frequency conversion) pumps $\Omega_k=k\Delta \ll \omega_0$ with $k\in\mathbb{N}$, and high frequency (HF, sometimes called squeezing) pumps $\Omega_{k'}=2\omega_0 + k'\Delta$ with $k'\in\mathbb{Z}$, and we decompose the  $2N \times 2N$ matrix $\matr{M}$ to a general form,
\begin{equation}\label{eq:Mexp}
    \matr{M} = \matr{M}_d + \sum_{k} \matr{L}_k + \sum_{k'} \matr{H}_{k'}.
\end{equation}
Here $\matr{M}_d$ is diagonal with elements $\Delta_m = \tfrac{1}{\gamma}(-\omega_m - \tilde{\omega}_0) \in \mathbb{C}$, and $\matr{L}_k$ and $\matr{H}_{k'}$ are the coupling matrices associated with the low- and high-frequency pumps, respectively. 

The matrices $\matr{L}_k$ exhibit two off-diagonals at $\pm 2k$,
\begin{equation}\label{eq:Mk_LF}
    \matr{L}_k = \begin{pmatrix}
        0 & & & l_k & & & & & \\[5 pt]
        & 0 & & & -l_k^* & & & & \\
         & & \ddots & & & \ddots & & & \\
         l_k^* & & & 0 & & & l_k & & \\[5 pt]
         & -l_k & & & 0 & & & -l_k^* \\
         & & \ddots & & & \ddots & & \\
         & & & l_k^* & & & 0 & \\[5 pt]
         & & & & -l_k & & & 0 \\
    \end{pmatrix},
\end{equation}
while the matrices $\matr{H}_{k'}$ contain a single anti-diagonal at $2k'$ 
\begin{equation}\label{eq:Mk_HF}
    \matr{H}_{k^\prime} = \begin{pmatrix}
             0 & & & & & & h_{k^\prime} & & \\[3 pt]
             & 0 & & & & -h_{k^\prime}^* & & & \\[-2 pt]
             & & \ddots & & \iddots & & & & \\
             & & \iddots & & \ddots & & & & \\
             & h_{k^\prime} & & & & 0 & & & \\[3 pt]
             -h_{k^\prime}^* & & & & & & 0 & & \\[-4 pt]
             & & & & & & & \ddots & \\[-2 pt]
             & & & & & & & & 0 \\
        \end{pmatrix}.
\end{equation}
For clarity, we distinguish between $l_k=\tfrac{g_k}{\gamma}$ and $h_{k'}=\tfrac{g_{k'}}{\gamma}$, which are the normalized complex pump amplitudes of the low- and high-frequency pumps, respectively.
There are $N_{LF}=N-1$ low frequency pumps and $N_{HF}=2N-1$ high-frequency pumps.
Thus, in total there are $N_{tot} = N_{LF} \, + \, N_{HF} = 3N-2$ pumps.

The solution of~\eqref{eq:EOM_matrix} is readily available upon matrix inversion, $ \vec{a} = \frac{i}{\sqrt{\gamma}} \matr{M}^{-1} \vec{a}_\text{in}$. 
The scattering matrix of input-output theory in the frequency domain, is then derived from the mode matching condition $\sqrt{\gamma}\vec{a} = \vec{a}_{\text{in}} + \vec{a}_{\text{out}}$, 
\begin{equation} \label{eq:S}
   \matr{S} = i\matr{M}^{-1} - \identity.
\end{equation} 
Multiple pumps contribute to non-zero off-diagonal elements in $\matr{M}$. Upon matrix inversion, these off-diagonal elements mix, creating high order intermodulation products appearing as off-diagonal elements of $\matr{S}$.
Note that these higher-order processes have different origin than those generated from higher order terms (e.g. Kerr)in $\mathcal{H}_{\text{PO}}$.

In summary, frequency-domain knowledge of the periodic pump allows for discrete convolution leading to $\matr{M}$ in \eqref{eq:EOM_matrix}, which upon inversion yields $\matr{S}$ in \eqref{eq:S}. 
Note that $\matr{S}$ in the $a_i, a_i^*$ basis is a $2N \times 2N$ complex matrix, with elements $S_{mn}=|S_{mn}|e^{i\phi_{mn}}$ defined by magnitude and phase, both determined by the combined action of the pumps.
This formally defines the direct problem illustrated in Fig.~\ref{fig:PO-setup}(c). 
The inverse parametric problem is the computation of the pump waveform necessary to realize a given target scattering matrix $\matr{S}_\odot$. 
Its solution requires de-convolution, which is generally not trivial. 

\subsection{Solving the inverse problem}

In the decomposition of $\matr{M}$ in~\eqref{eq:Mexp}, the pump-induced coupling matrices $\matr{L}_k$, and $\matr{H}_{k'}$ exhibit internal symmetries, reflecting the identical action of each pump tone across all frequency modes.  
These symmetries allow us to further decompose the coupling matrices as 
\begin{equation}
    \matr{L}_k = l_k\,\matr{L}_k^{+} + l_k^{*}\,\matr{L}_k^{-}, 
    \qquad
    \matr{H}_{k'} = h_{k'}\,\matr{H}_{k'}^{+} + h_{k'}^{*}\,\matr{H}_{k'}^{-},
    \label{eq:Lk_Hk_decomp}
\end{equation}
where the $l_k,\,h_{k'} \in \mathbb{C}$ are normalized complex pump amplitudes, and the matrices $\matr{L}_k^{\pm}$ and $\matr{H}_{k'}^{\pm}$, which have only elements $\pm1$ and $0$, are defined in Appendix~\ref{appA}.

To determine the complex pump amplitudes $l_k$ and $h_{k'}$ we introduce the Frobenius (Hilbert–Schmidt) inner product between two matrices,
\begin{equation}\label{eq:Frob}
    (\matr{A},\matr{B}) = \text{Tr}(\matr{A}\cdot \matr{B}^T),
\end{equation}
With respect to this inner product the set of all $\matr{L}_k^{\pm}$ and $\matr{H}_{k'}^{\pm}$ form a orthogonal basis in the space of complex $2N\times2N$ matrices:
\begin{align}
    (\matr{L}_k^{\pm}, \matr{L}_q^{\pm}) &= 2(N-k)\,\delta_{kq},  \nonumber\\
    (\matr{H}_{k'}^{\pm}, \matr{H}_{q'}^{\pm}) &= (N-|k'|)\,\delta_{k'q'},  \nonumber\\
    (\matr{L}_k^{\pm}, \matr{H}_{q'}^{\pm}) &= 0. \label{eq:LH_orthogonality}
\end{align}

We define the off-diagonal components of $\matr{M}$ as 
$\Delta \matr{M} \equiv \matr{M} - \matr{M}_d$, 
which collects all pump-induced mode couplings. 
Eq.~\eqref{eq:Mexp} together with Eq.~\eqref{eq:Lk_Hk_decomp}, and the orthogonality relations~\eqref{eq:LH_orthogonality}, imply that we can expand $\Delta \matr{M}$ as
\begin{equation}
    \Delta \matr{M} =
    \sum_{k}\!\left(l_k\,\matr{L}_k^{+} + l_k^{*}\,\matr{L}_k^{-}\right)
    + \sum_{k'}\!\left(h_{k'}\,\matr{H}_{k'}^{+} + h_{k'}^{*}\,\matr{H}_{k'}^{-}\right).
    \label{eq:M_expansion}
\end{equation}
Thus the matrices $\{\matr{L}_k^{\pm},\matr{H}_{k'}^{\pm}\}$ constitute an orthogonal basis spanning the Hilbert space that contains all possible coupling matrices $\Delta\matr{M}$. 
In this framework, the complex pump amplitudes are expansion coefficients, obtained by projecting $\Delta \matr{M}$ onto the corresponding basis elements:
\begin{equation}
    l_k^{(*)} = 
    \frac{(\Delta \matr{M}, \matr{L}_k^{\pm})}
    {\|\matr{L}_k^{\pm}\|}, 
    \qquad 
    h_{k'}^{(*)} = 
    \frac{(\Delta \matr{M}, \matr{H}_{k'}^{\pm})}
    {\|\matr{H}_{k'}^{\pm}\|},
    \label{eq:projection}
\end{equation}
where $\|\matr{A}\| = \sqrt{(\matr{A},\matr{A})}$ denotes the Frobenius norm.

In summary, given a target scattering matrix $\matr{S}_{\odot}$, Eq.~\eqref{eq:S} is inverted to yield the corresponding target equation-of-motion matrix $\matr{M}_{\odot}$, and its off-diagonal components $\Delta \matr{M}_{\odot}$:
\begin{equation}
     \Delta \matr{M}_{\odot} = -i\,(\matr{S}_{\odot} + \mathbb{1})^{-1} - \matr{M}_{\odot \,d}.
    \label{eq:M_target}
\end{equation}
Substituting $\Delta\matr{M}_{\odot}$ into Eqs.~\eqref{eq:projection} provides the complete set of complex coefficients $\{l_k, h_{k'}\}$, directly yielding the amplitude and phase of each pump tone.  
This method, which we refer to as the \emph{Pump Projection Method} (PPM), constitutes an exact and closed-form solution to the inverse parametric problem. The complexity of the method scales linearly with the total number of pumps which is $O(N)$ in the number of modes.

\subsection{Testing the Pump Projection Method}

We validate the PPM through numerical simulations on a set of $10^4$ randomly generated target scattering matrices of different sizes $2N$.
Each target matrix $\matr{S}_{\odot}$ is produced by solving the direct problem for a randomly generated pump waveform consisting of $3N-2$ tones, both low and high frequency pumps with random amplitude and phase (see appendix \ref{appA}).  
Solving the inverse problem for $\matr{S}_\odot$ yields the corresponding pump waveform, which we use in the direct problem to reconstruct a scattering matrix $\matr{S}_\text{rec}$. 
The deviation between $\matr{\matr{S}_{\odot}}$ and $\matr{S}_\text{rec}$ benchmarks the accuracy of the method.
Up to numerical precision, our implementation of the PPM is always able to recover the original pump waveform.
\begin{figure}
    \centering
    \includegraphics[width=\columnwidth]{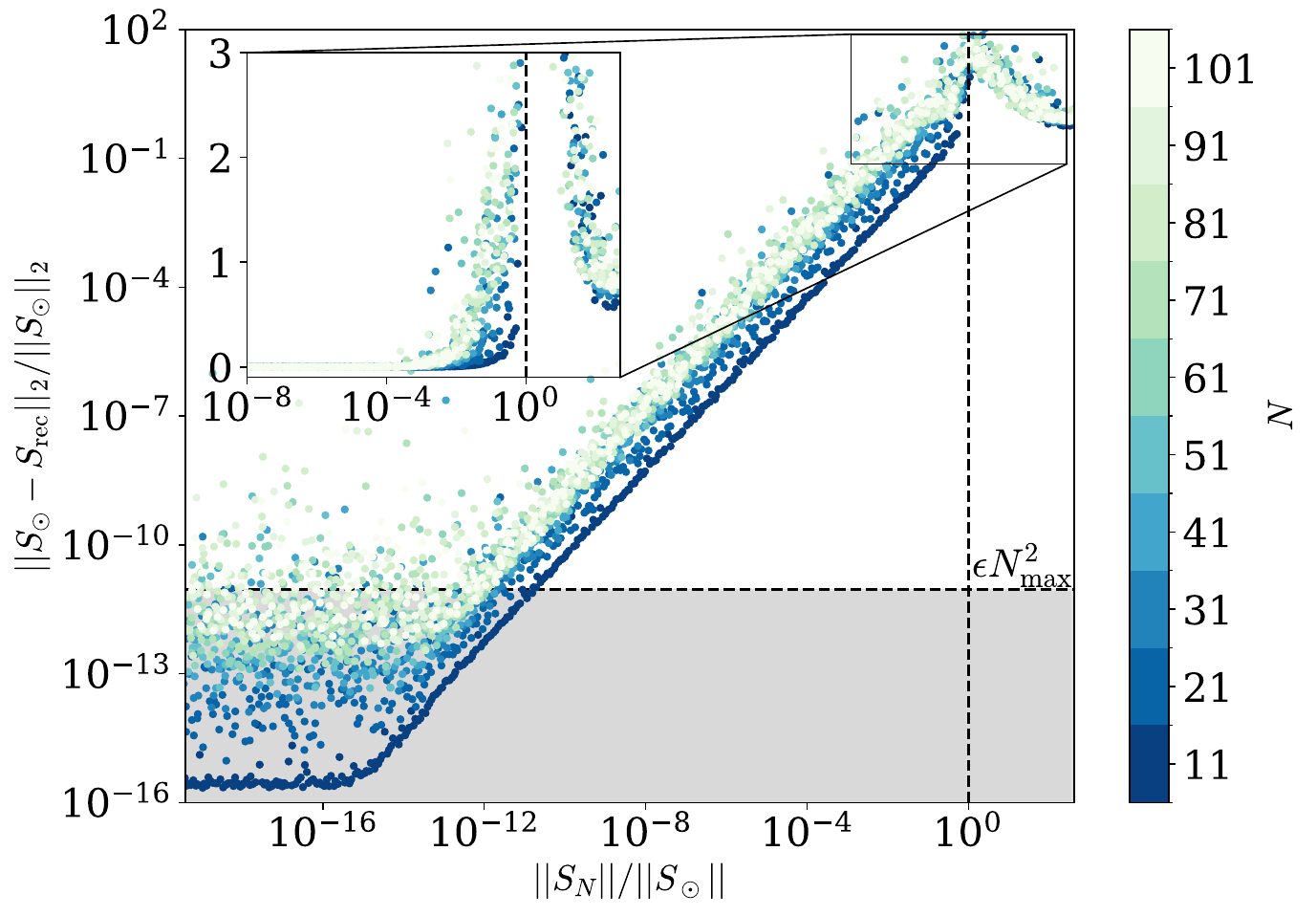}
    \caption{Robustness of the PPM against Gaussian noise. Relative reconstruction error $\| \matr{S}_\odot - \matr{S}_\text{rec} \|_2 / \| \matr{S}_\odot \|_2$ as a function of the noise-to-target ratio $\| \matr{S}_N \|_2 / \| \matr{S}_{\odot} \|_2$ for different matrix sizes $2N$.  
    The horizontal dotted line indicates the numerical-precision for the maximum number of modes tested $N_\text{max}$. 
    The vertical dotted line marks unity noise-to-target ratio, beyond which the PPM becomes unreliable.}
    \label{fig:Error_S}
\end{figure}

\begin{figure*}
    \centering
    \includegraphics[width=\textwidth]{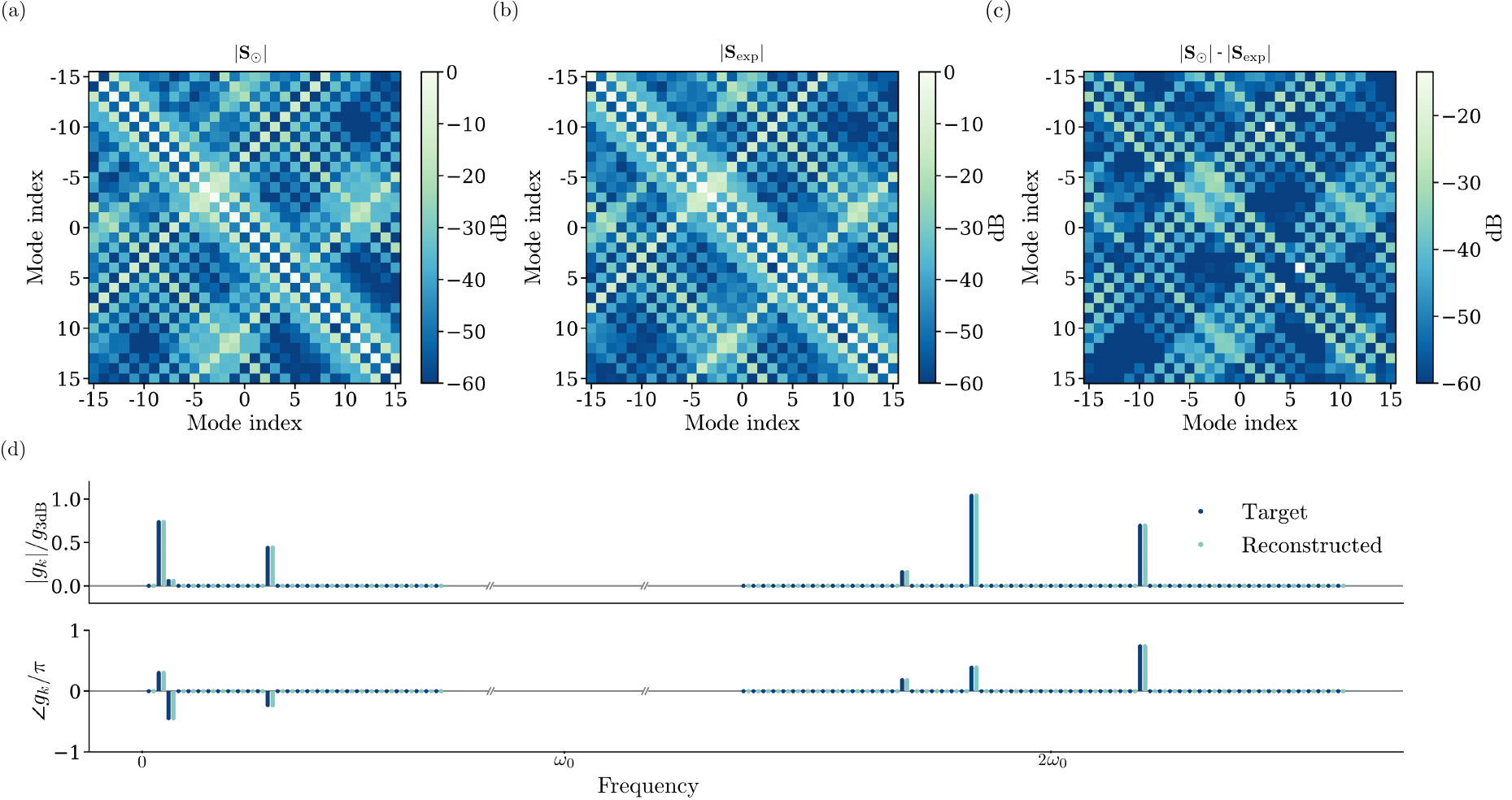}
    \caption{Experimental validation of the Pump Projection Method (PPM). 
    (a) Target scattering matrix $\matr{S}_{\odot}$ constructed from the target pumps (blue) shown in (d). 
    (b) Experimentally measured scattering matrix $\matr{S}_{\mathrm{exp}}$ obtained from the reconstructed pumps (cyan) shown in (d).
    (c) Difference between the magnitudes of the target and measured scattering matrices.
    (d) Pump amplitude (upper) and phase (lower). Amplitudes are normalized by the single-frequency pump amplitude at $2\omega_0$, which achieves 3dB gain in the JPA.}
    \label{fig:PPM_experiment}
\end{figure*}

To assess robustness, Gaussian noise of variable magnitude is added to the target matrix $\matr{S}_\odot$. 
We quantify the error in the PPM as the relative deviation from the noise-free target $\| \matr{S}_\odot - \matr{S}_{\mathrm{rec}} \|_2 / \| \matr{S}_\odot \|_2$ where $\| \cdot \|_2$ denotes the matrix 2-norm.  In Fig.~\ref{fig:Error_S} we plot this relative deviation as a function of the noise-to-target ratio $\|\matr{S}_N\|_2/ \|\matr{S}_\odot \|_2$. 
We averaged over 100 noise configurations and repeated the study for different numbers of modes $N$, corresponding to different matrix sizes. 
As shown in Fig.~\ref{fig:Error_S}, the error scales linearly with the noise magnitude and remains bounded from below by numerical precision $\epsilon$.
The grey shaded area marks the region dominated by numerical noise $\epsilon N^2$, for the largest matrix size $2N_\text{max}$. 
The inset illustrates the divergence of the error as the noise-to-target ratio approaches unity, indicating that the method breaks down.
Thus, the PPM gives a faithful reconstruction of the pump waveform, even in the presence of significant noise, across all matrix sizes.

In addition to numerical tests, we verify the performance of the PPM with an experiment on a Josephson Parametric Amplifier (JPA) in a setup described in Appendix~\ref{appB}. 
Figure~\ref{fig:PPM_experiment} shows a representative example in which the PPM reconstructs the pump waveform for a given target scattering matrix. 
Panel~\ref{fig:PPM_experiment}(a) displays the target matrix $\matr{S}_{\odot}$ obtained from a randomly generated pump waveform, and panel~\ref{fig:PPM_experiment}(b) shows the experimentally measured matrix $\matr{S}_{\mathrm{exp}}$ obtained from the reconstructed pump waveform. 
Here, the mode index $0$ corresponds to the resonant frequency of the JPA.
The two matrices exhibit excellent agreement as shown in Figure~\ref{fig:PPM_experiment}(c), with a total relative deviation of $\| \matr{S}_\odot - \matr{S}_{\mathrm{exp}} \|_2 / \| \matr{S}_\odot \|_2 = 0.037$; the residual discrepancy is primarily set by the experimental background noise floor in Fig.~\ref{fig:PPM_experiment}(b). 
Panel~\ref{fig:PPM_experiment}(d) shows the corresponding pump waveforms, illustrating the distribution of amplitudes and phases for the target pump tones (blue) and reconstructed (cyan). 
These results validate the PPM beyond numerical simulation, demonstrating its application in an experiment.

\subsection{Applications}

The PPM is very effective at recovering the pumping scheme when one has full knowledge of both magnitude and phase of the elements of the target scattering matrix.
\begin{figure*}
    \centering
    \includegraphics[width=\textwidth]{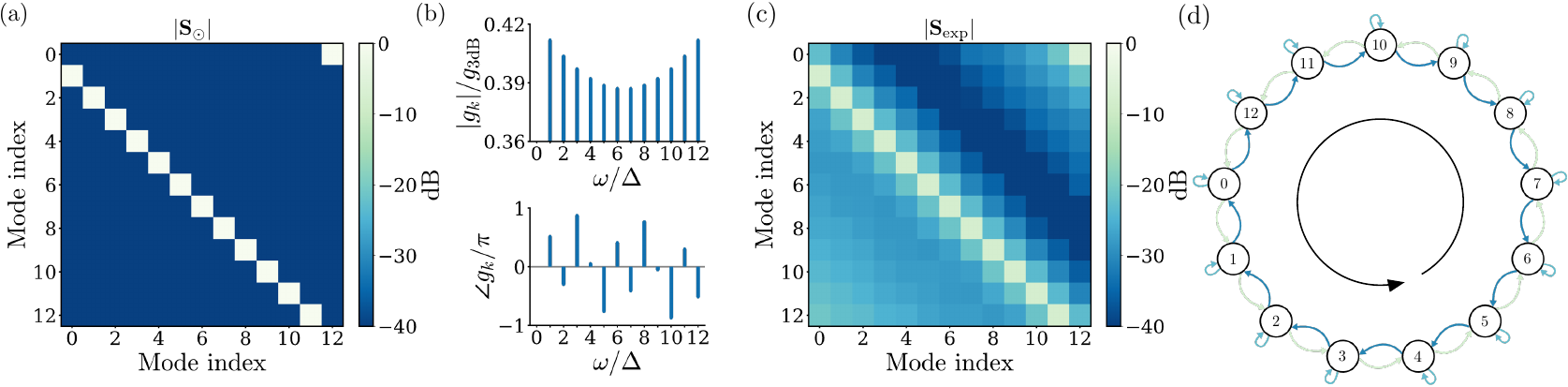}
    \caption{Design and implementation of a 13-mode circulator using the PPM. 
    (a) Target scattering matrix $|\matr{S}_\odot|$ designed in the $xp$ basis. 
    (b) Low frequency pump amplitudes (top) and phases (bottom) of the waveform returned by the PPM. 
    (c) Experimentally measured scattering matrix $|\matr{S}_{\text{exp}}|$, obtained by applying the pump waveform from panel (b). 
    (d) Graphical representation of the 13-mode circulator. Black arrow shows the circulation direction.}
    \label{fig:circulator_experiment}
\end{figure*}
A separate but equally significant issue is the design of a target scattering matrix which is physically realizable.  When designing a target scattering matrix for a practical application, e.g. a signal routing  between frequency modes, we often consider only magnitudes, which define how power is distributed between modes.
The associated phases may be unknown, yet they may play a significant role in the redistribution of power.  
For a given set of target magnitudes, only specific phase configurations correspond to physically realizable scattering matrices. 
The PPM projects onto a pump configuration which generates a closely associated, but realizable scattering matrix. 
Consequently, if the phases of the target are inconsistent with the desired magnitudes, the recovered pumping scheme generates scattering which may deviate significantly from the target.
This phase ambiguity of $\matr{S}$ is a known problem in literature~\cite{devaney_nonuniqueness_1978, defacio_nonuniqueness_1992, dersy_reconstructing_2024, del_hougne_virtual_2025}. Causality between the input and the output enforce constraint on the real and imaginary part of $\matr{S}$ effectively imposing specific phase relations on the scattering matrix elements~\cite{kronig_supplementary_1946, schutzer_connection_1951, kampen_causalite_1961}. 

We know of no universal or optimal strategy to construct a good scattering matrix, but we have identified a few physical constraints and practical guidelines:
(i) It is often easier to start in the $(x_i,p_i)$ basis with $\matrx{S_\odot}=U\matr{S_\odot} U^\dag$, where $U$ is the canonical transformation from $(a_i,a_i^\dagger)$ to $(x_i,p_i)$. In the $(x_i,p_i)$ basis we can apply the constraint that $\matrx{S_\odot}$ must be both real and symplectic, automatically enforcing the correct phase relations between creation and annihilation operators.
(ii) Symmetry considerations, e.g non-reciprocity, can serve as valuable criteria for guiding the design.
(iii) Enforcing causality by imposing the Kramers-Kronig relations between the real and imaginary elements of the target scattering matrix. 
(iv) An iterative workflow, where the target scattering matrix is progressively refined. A new target is constructed by altering a calculated scattering matrix using pumps returned by the PPM in the previous iteration.

Nevertheless, we used the PPM to construct a complicated pump waveform for $13$-mode circulation, which obviously satisfies the above criteria because we verified it in an actual experiment, as shown in Fig.~\ref{fig:circulator_experiment}. 
The nonreciprocal target scattering matrix of an ideal 13-mode circulator is shown in Fig.~\ref{fig:circulator_experiment}(a).  
This ideal scattering matrix is not physically realizable for arbitrary magnitude of the off diagonal elements. 
Following the principles discussed above we use the PPM to iterate through the magnitude of the off-diagonal elements of $\matr{S}_\odot$, maximizing for nonreciprocity of the recovered scattering matrix, $\max\left[|\matr{S}_\text{rec}|-|\matr{S}^T_\text{rec}|\right]$. 
We arrive at the pumping scheme shown in Fig.~\ref{fig:circulator_experiment}(b) which contains only LF pumps.
Programming this pumping scheme into our multifrequency lock in amplifier, we generate the measured scattering matrix shown in Fig.~\ref{fig:circulator_experiment}(c).  Panel~\ref{fig:circulator_experiment}(d) illustrates the resulting circular routing between the 13 frequency modes, where the color coded arrows depict the magnitude of the dominant scattering processes. 
We do not achieve ideal circulation, but the unwanted scattering channels are suppressed by more than -20~dB, with approximately -6~dB transmission between modes in the desired direction of circulation.

\subsection{Dynamic control of Scattering}

The PPM is useful for determining pumping schemes to realize complicated scattering matrices, but lack of knowledge of the phases of $\matr{S}$ define the need for alternative approximation methods which are fast and easy to implement.
Here we demonstrate a method to quickly calculate a pumping scheme which approximates the desired scattering, when only the magnitude is known.
We can rapidly reprogram the pumps in an experiment where the pump waveform is updated every $T_\text{col}=50$~ms. 
Each pump configuration is designed to generate a matrix  $\matr{S}_{\ell}$ which scatters a single input tone into a desired set of output tones that code the pixel intensities of the $\ell^\text{th}$ column of an image.
To a good approximation the pump waveform that realizes each $\matr{S}_{\ell}$ is composed of $N_p$ HF pumps at the frequencies $\Omega_k = 2\omega_0 + k\Delta$ where $k\in[-N_p,0]$, with amplitude $|h_k|\propto \text{Pix}_{k\ell}$, and with random phase $\phi_k=\text{rand}[0, 2\pi]$. 
The choice of random phase facilitates average cancellation of the higher order mixing processes through destructive interference, allowing the second-order intermodulation products generated by the HF pumps to dominantly contribute to the target $\matr{S}_{\ell}$. 

\begin{figure}
    \centering
    \includegraphics[width=\columnwidth]{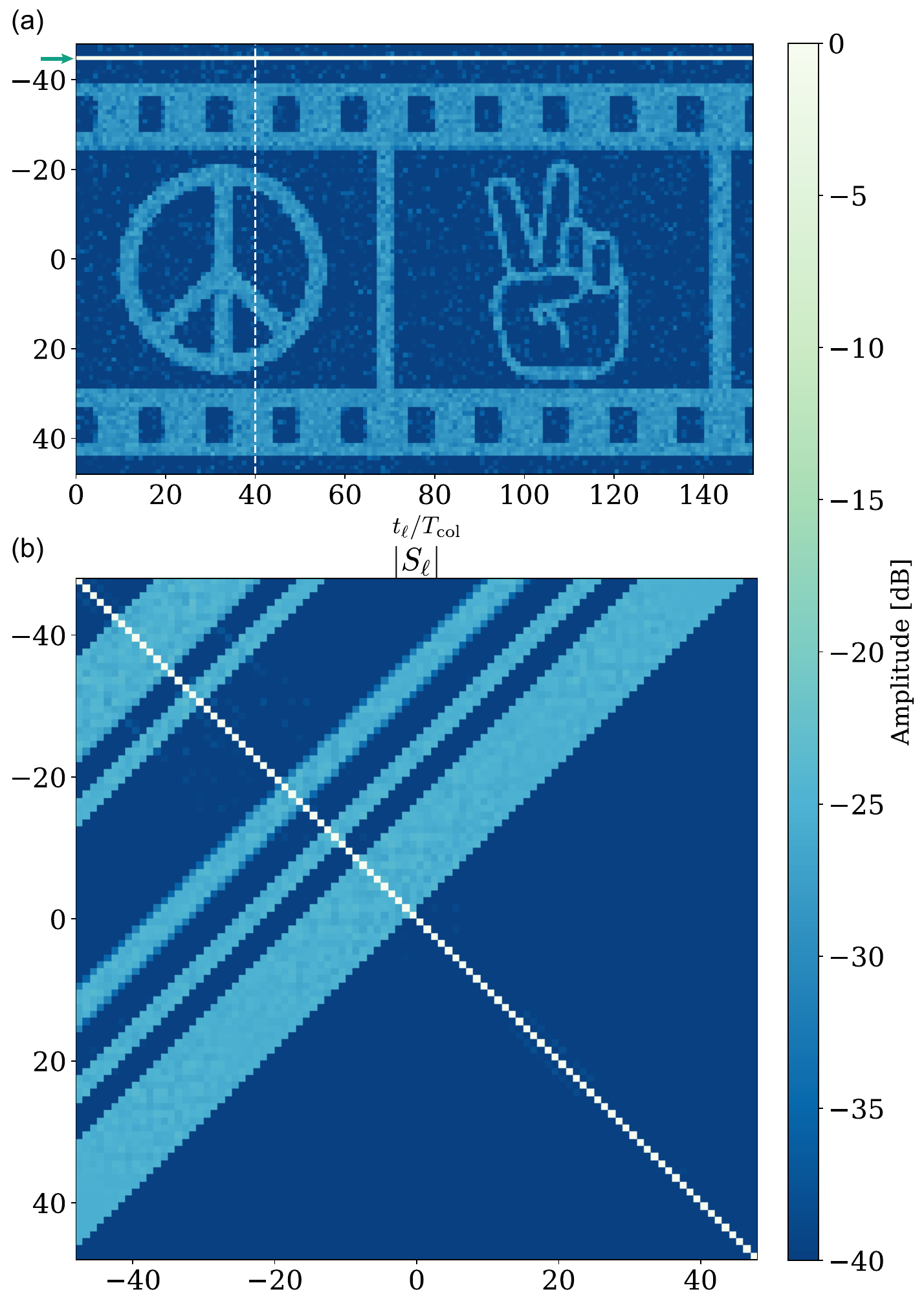}
    \caption{(a) Measured output mode amplitudes as a function of time $t_\ell/T$. 
    The pump waveform is dynamically updated every measurement window $T$, encoding successive columns of the target image. 
    The scattered output evolves in time to reproduce the desired pattern, with amplitudes expressed in decibels and normalized to the pump-off reference.
    (b) Example of target scattering matrix $\matr{S}_\ell$ used to generate the output at $t_\ell/T=40$ (white dashed line in panel (a)).
    }
    \label{fig:Image_Scattered}
\end{figure}

We performed an experiment encoding an image which is reproduced in Fig.~\ref{fig:Image_Scattered}(a).  
An example scattering matrix $\matr{S}_{\ell}$ used to reproduce the column of pixels at $t_\ell/T=40$ (white dashed line) is shown in \ref{fig:Image_Scattered}(b). 
During each time window starting at $t_\ell=\ell\,T_\text{col}$, $\ell\in[1,N_p]$: (i) a weak coherent tone is continuously applied at a fixed input frequency $\omega_{-45}$ (green arrow in Fig.~\ref{fig:Image_Scattered}(a)); (ii) the pump waveform corresponding to the $\ell$-th image column is applied to the flux port of the JPA; and (iii) the full set of output frequency modes is simultaneously measured using the  multifrequency lock-in amplifier.  

Figure~\ref{fig:Image_Scattered} shows the measured amplitudes of the output modes as a function of normalized time $t/T_\text{col}$.
The scattered output evolves dynamically over the sequence, accurately reproducing the image encoded in the time series of pump waveforms. 
The amplitudes are expressed in decibels and normalized to the pump-off reference case. 
The high contrast in the image generated by rapidly changing the scattering of a single tone,  highlights the ability to realize arbitrary, time-dependent transformations of the scattering matrix. 
See Appendix~\ref{appB} for implementation details.

\subsection{Gaussian State Engineering}

The ability to engineer the scattering matrix is of particular significance in continuous-variable quantum information~\cite{weedbrook_gaussian_2012} with Gaussian states, which are completely characterized by their first and second statistical moments.
For such states the scattering matrix acts as a Gaussian transformation on the covariance matrix 
\begin{equation}
    \matr{V}=\matrx{S}\matr{V}_\text{in}\matrx{S}^T,
\end{equation}
where $\matr{V}_\text{in}$ denotes the input covariance matrix. 
Throughout this section we use $\matrx{S}$ to denote the scattering matrix in the quadrature basis $\vec{x}=(x_1,p_1,x_2,p_2,\dots,x_N,p_N)$, related to its representation in the annihilation-operator basis by $\matrx{S} = \matr{U}\matr{S}\matr{U}^\dag$, where $\matr{U}$ is the canonical transformation between the two bases.
In this representation, $\matrx{S}$ is real and symplectic.

If vacuum is injected into the PO, the input covariance matrix is $\matr{V}_\text{in}=\tfrac{\hbar}{2}\mathbb{1}$ and the output covariance matrix is determined by the scattering matrix
\begin{equation}
    \matr{V}=\frac{\hbar}{2}\matrx{S}\matrx{S}^T. 
    \label{S_to_V}
\end{equation}
Equation~\eqref{S_to_V} connects a desired Gaussian state with the symplectic transformation that generates it.

There is no unique solution to the problem of finding a scattering matrix $\matrx{S}$ that satisfies~\eqref{S_to_V} for a given covariance matrix $\matr{V}$.
Any orthogonal matrix $\matr{Q}$ ($\matr{Q}\matr{Q}^T=\mathbb{1}$) generates a matrix $\matrx{S}^\prime = \matrx{S}\matr{Q}$ that also satisfies Eq.~\eqref{S_to_V}.
Among the infinitely many solutions, we are interested in those that can be generated by pumping a parametric oscillator with a periodic waveform which is retrievable with the PPM.
Typical decompositions of $\matrx{S}\matrx{S}^T$, such as Wigner-Moyal or Cholesky ~\cite{simon_gaussian_1988,menicucci_graphical_2011} tend to produce strongly asymmetric and non-reciprocal symplectic matrices that are typically difficult or impractical to realize through parametric processes.

A particularly relevant subclass is that of \emph{symmetric} scattering matrices, $\matrx{S}=\matrx{S}^T$, used in many key applications such as multimode squeezed states and graph-like continuous-variable cluster states.
For this class $\matr{V}=\tfrac{\hbar}{2}\matrx{S}^2$, and the problem can be inverted by computing the matrix square-root 
\begin{equation}
    \matrx{S}=\sqrt{_M \matr{V}}   \label{sqrtV}
\end{equation}
where the symbol $\sqrt{_M \;\cdot}$ denotes the matrix square-root (see Appendix~\ref{app_Msqrt} for details).
Equation~\eqref{sqrtV} retrieves a scattering matrix $\matr{S}_\odot$, given a target covariance matrix $\matr{V}_\odot$.
With this target scattering matrix $\matr{S}_\odot$ the projection method allows us to obtain a pump waveform to realize the Gaussian state described by $\matr{V}_\odot$.

\section{\label{sec:conclusion}Conclusion}

We introduced the Pump Projection Method (PPM), a general and closed-form solution to the inverse parametric problem for a single pole parametric circuit. 
By projecting the off-diagonal structure of the equation-of-motion matrix onto an orthogonal basis, the PPM directly returns the complete set of pump amplitudes and phases required to realize a desired scattering matrix.
We validated the method numerically and studied its sensitivity to the addition of controlled Gaussian noise. 
The reconstruction error scales linearly with the noise amplitude and remains bounded by numerical precision, demonstrating robustness of the method.

We tested the method in experiments with a Josephson Parametric Amplifier (JPA) thermalized at 10~mK. 
With a randomly generated target scattering matrix, we showed that the method recovers a pump waveform that, when applied to the JPA , realizes a measured scattering matrix with a relative deviation to the target of $3.7\%$. 
We used the PPM to engineer circulation between $13$ frequency modes in the JPA, with more than -20~dB attenuation of scattering to unwanted channels, and transmission of -6~dB in the desired direction. 
These results demonstrate that the PPM is a practical method which reliably enables the design and implementation of complex scattering that is extremely challenging to obtain by intuition alone. 
However, the PPM will not work if the target is not physically realizable. The applicability of the method requires the design of a target scattering matrix that satisfies simplecticity and causality constraints. This may require additional optimization techniques if only partial knowledge of the desired target is available (e.g. only amplitudes $|S_{\odot\,mn}|$ are known).
We further introduce an approximate method for designing a pump that is capable of rapidly reconfiguring scattering, to dynamically encode arbitrary information in the output modes.

An interesting extension of the PPM is the exploration of frequency-mode coupling by parametrically modulating circuits with multiple poles.  
For example a parametric band-pass filter which sharply defines region of the spectrum where frequency modes scatter in to one-another.
The ability to engineer the scattering matrix is of particular importance for continuous-variable quantum information~\cite{weedbrook_gaussian_2012} where the primary resources are Gaussian states which are fully specified by their first and second moments. 
The PPM provides a direct way to compute the pump waveform and realize a desired covariance matrix. 
Its application with a multifrequency lockin amplifier makes for easy experimental realization of very complex mode couplings, opening up new opportunities for Gaussian-state engineering and multi-modal quantum technologies.

\begin{acknowledgments}
We acknowledge Joe Aumentado at the National Institute of Standards and Technology (NIST) for helpful discussions and for providing the JPA used in this experiment. 
This work was partially supported by the Knut and Alice Wallenberg Foundation through the Wallenberg Center for Quantum Technology (WACQT), the Swedish Natural Science Research Council (VR) and the Olle Engkvist foundation.
\end{acknowledgments}

\section*{Author Declaration}

F. L. derived the theoretical method, M. C. implemented it numerically. F. L. and M. C. performed the numerical benchmarking and the validating experiments. D. B. H. supervised the work. All authors edited, discussed and reviewed the manuscript.

D. B. H. is part owner of the company Intermodulation Products AB, which produces the digital microwave platform used in this experiment.

\section*{Data Availability}
The data that support the findings of this study are openly available in Zenodo at~\cite{lingua_data_2026}.

\appendix

\section{Experimental setup}\label{appB}

\begin{figure}
    \centering
    \includegraphics[width=\columnwidth]{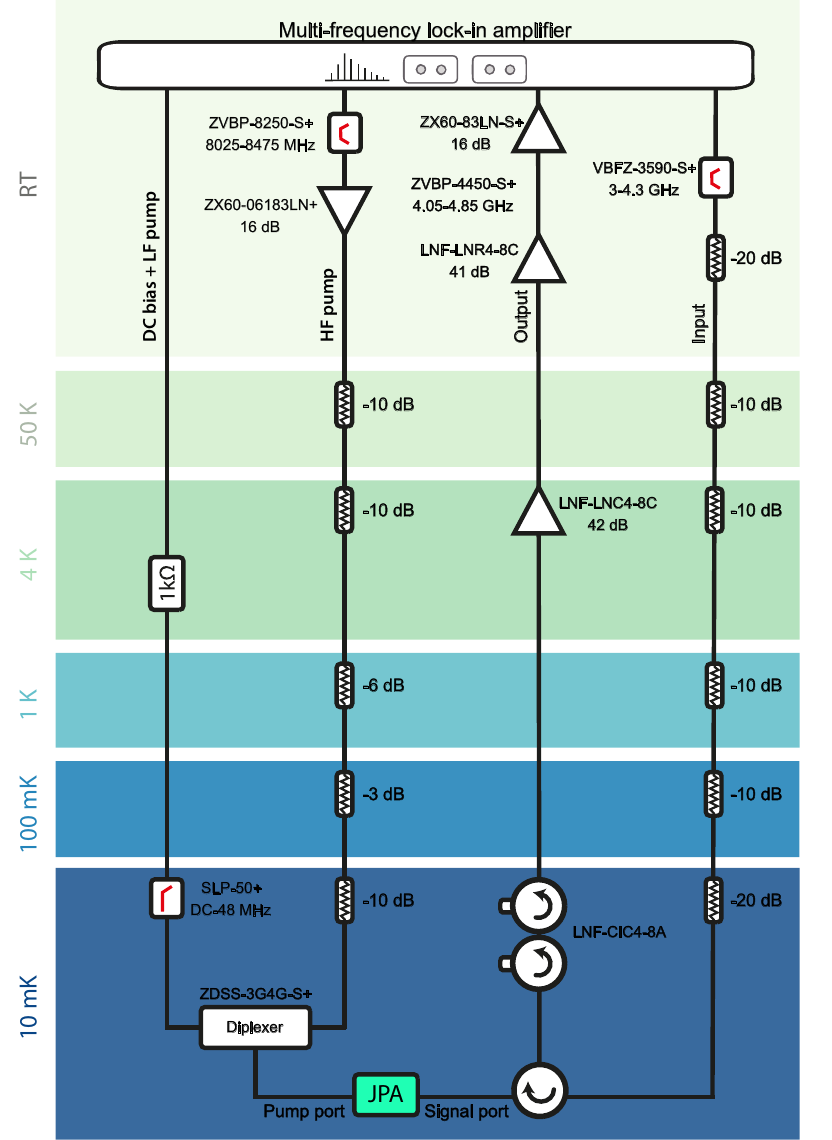}
    \caption{Experimental setup used in this paper with a Josephson Parametric Amplifier used as the parametric oscillator.
    }
    \label{fig:setup}
\end{figure}

Figure \ref{fig:setup} shows a schematic of the experimental setup.  The experiments are performed in a dilution refrigerator (Bluefors LD250) at millikelvin temperature. 
The parametric oscillator is Josephson Parametric Amplifier (JPA) designed for broad-band (100 MHz) gain, with loaded quality factor of about $Q = 37.5$. 
The JPA resonance is tuned by magnetic flux threading a SQUID loop.  
The DC flux adjusts the operating point (resonance frequency) and the AC flux is externally controlled by the pump waveform.
DC bias and low-frequency pumps are combined with high frequency pumps using a cryogenic bias-tee. 
The input and output share the same signal port and they are separated close to the JPA by a cryogenic circulator. 
The reflected signal passes through two isolators before reaching a low-noise HEMT amplifier at $4~\mathrm{K}$. 
Attenuation and filtering are distributed along the signal line to thermalize the incoming radiation and suppress noise from higher temperature stages.

A typical experiment uses a DC flux close to half a flux quantum, where the resonance frequency is approximately $4.2~\mathrm{GHz}$. 
The pumps modulates the SQUID inductance, generating frequency mixing among the modes connected to the signal port.
All pump and probe tones are synthesized and detected by a digital multi-frequency lock-in amplifier (PRESTO, Intermodulation Products AB).
The instrument performs coherent modulation and demodulation at up to 192 frequencies simultaneously (i.e. in the same time window) with one reference phase for all frequencies, and digital control of the amplitude and phase of both signal and pump. 
Scattering measurements typically average over $5\times10^3$ windows $T$. 
The second Nyquist zone of the digital-to-analog converters is used to generate microwave signals from 3 to 10~GHz.  
Under-sampling and aliasing are used to down-convert the measured signals to the first Nyquist zone of the analog-to-digital converters. 
Analog filters at room temperature select the desired Nyquist zones for both up- and down-conversion.
With this approach, digital multipliers replace the analog IQ mixers traditionally used at microwave frequencies.

This all-digital approach enables phase coherence between all frequency components, essentially defining the orthonormal comb of modes in the measurement time window $T$, and the corresponding frequency spacing of the modes $\Delta = 1/T$. 
All frequencies in the comb and the digital sampling clock frequency are related by a rational fraction, establishing a frequency-comb basis, ensuring orthogonality of the modes and eliminating Fourier leakage between them. 
In the measurements presented here, the spacing was set to $\Delta = 100~\mathrm{kHz}$, much smaller than the linewidth of the JPA $\sim 100~\mathrm{MHz}$, so that all modes of the comb are well within the amplifier bandwidth.
Specifically, the 13-mode circulator of Fig.~4 spans a total bandwidth of $12\Delta = 1.2$~MHz, while the dynamic scattering experiment of Fig.~5 employs 97 frequency modes covering a bandwidth of $96\Delta = 9.6$~MHz, both centered around the JPA resonance frequency $\omega_0/2\pi \approx 4.2$~GHz.
One limitation of digital synthesis is the dynamic range of the multifrequncy pump waveform, where the ratio of the largest to the smallest Fourier coefficient cannot exceed the resolution of the 14-bit digital-to-analog converters, i.e. $2^{14} = 16384 $.

Prior to each measurement, the JPA is characterized by measuring its flux modulation curve with a vector network analyzer (VNA), allowing us to identify the DC flux bias corresponding to the resonance frequency $\omega_0/2\pi \approx 4.2$~GHz and to ensure operation in the linear regime. 
The HF and LF pump amplitudes are calibrated independently, as they are delivered through separate lines with different attenuation. 
The HF pump amplitudes are referenced to a single pump tone at $2\omega_0$ producing 3~dB of gain. 
The LF pump amplitudes are calibrated by matching the amplitude of an idler generated by a single LF pump, to those produced by the reference HF pump. This calibration gives a conversion factor that is programmed into the measurement software.

\section{The coupling matrices} \label{appA}

The pump coupling matrices ~\eqref{eq:Mk_LF} and~\eqref{eq:Mk_HF} contain the collective action of each pump tone on the different frequency modes. Their structure is revealed by extracting the constants $l_k$, $l_k^*$ and $h_{k^\prime}$, $h_{k^\prime}^*$. 
\begin{equation}
    \matr{L}_k = l_k \matr{L}_k^+ + l_k^* \matr{L}_k^-,
\end{equation}
\begin{equation}\label{eq:Mk_exp}
    \matr{H}_{k^\prime} = h_{k^\prime} \matr{H}_{k^\prime}^+ + h_{k^\prime}^* \matr{H}_{k^\prime}^-,
\end{equation}
$\matr{L}_k^\pm$, and $\matr{H}_{k^\prime}^\pm$ are constant, $2N\times 2N$ matrices of the form:
\begin{equation}\label{app_Gkp}
    \matr{L}_k^+ = \begin{pmatrix}
        \;\;0 & & & 1 & & & & & \\[5 pt]
        & \;\;0 & & & 0 & & & & \\
         & & \ddots & & & \ddots & & & \\
         \;\;0 & & & 0 & & & 1 & & \\[5 pt]
         & -1 & & & \;\;0 & & & \;\;0 \\
         & & \;\;\ddots & & & \;\;\ddots & & \\
         & & & 0 & & & 0 & \\[5 pt]
         & & & & \;\;-1 & & & \;\;0 \\
    \end{pmatrix},
\end{equation}
\begin{equation}\label{app_Gkm}
    \matr{L}_k^- = \begin{pmatrix}
        \;\;0 & & & 0 & & & & & \\[5 pt]
        & \;\;0 & & & -1 & & & & \\
         & & \ddots & & & \ddots & & & \\
         \;\;1 & & & 0 & & & 0 & & \\[5 pt]
         & 0 & & & \;\;0 & & & -1 \\
         & & \;\;\ddots & & & \;\;\ddots & & \\
         & & & 1 & & & 0 & \\[5 pt]
         & & & & \;\;0 & & & \;\;0 \\
    \end{pmatrix},
\end{equation}
\begin{equation}\label{app_Lkp}
    \matr{H}_{k^\prime}^+ = \begin{pmatrix}
             \;\;0 & & & & & & \;\;1 & & \\[3 pt]
             & \;\;0 & & & & \;\;0 & & & \\[-2 pt]
             & & \ddots & & \iddots & & & & \\
             & & \iddots & & \ddots & & & & \\
             & \;\;1 & & & & \;\;0 & & & \\[3 pt]
             \;\;0 & & & & & & \;\;0 & & \\[-4 pt]
             & & & & & & & \ddots & \\[-2 pt]
             & & & & & & & & 0 \\
        \end{pmatrix},
\end{equation}
\begin{equation}\label{app_Lkm}
    \matr{H}_{k^\prime}^- = \begin{pmatrix}
             0 & & & & & & 0 & & \\[3 pt]
             & 0 & & & & -1 & & & \\[-2 pt]
             & & \ddots & & \iddots & & & & \\
             & & \iddots & & \ddots & & & & \\
             & 0 & & & & 0 & & & \\[3 pt]
             -1 & & & & & & 0 & & \\[-4 pt]
             & & & & & & & \ddots & \\[-2 pt]
             & & & & & & & & 0 \\
        \end{pmatrix}.
\end{equation}

\newpage
\section{Matrix Square Root}\label{app_Msqrt}

Let us define the matrix square root $\sqrt{_M \;\cdot}$ as the inverse operation that allows to solve for $\matr{X}$ the matrix quadratic equation
\begin{equation}
    \matr{X}^2= \matr{B},   \label{Msquare}
\end{equation}
Let $\matr{W}$ be the transformation that diagonalizes $\matr{B}$, and $\matr{B}_d=\matr{W}\matr{B}\matr{W}^T$ the diagonal matrix containing the eigenvalues of $\matr{B}$.
Applying the transformation $\matr{W}$ to both sides of~\eqref{Msquare}
we obtain the quadratic relation on the diagonal matrix $\matr{B}_d$
\begin{equation}
    \matr{X}^\prime\matr{X}^\prime = \matr{B}_d, \label{Msquare_d}   
\end{equation}
where $\matr{X}^\prime=\matr{W}\matr{X}\matr{W}^T$.
For the properties of diagonal matrices, the solution of~\eqref{Msquare_d} is
\begin{equation}
    X^\prime=\begin{pmatrix}
        \pm\sqrt{b_1}&&&\\
        &\pm\sqrt{b_2}&&\\
        &&\ddots&\\
        &&&\pm\sqrt{b_N}\\
    \end{pmatrix}\label{d_sol}
\end{equation} 
where $b_{i}$ are the eigenvalues of $\matr{B}$.
From~\eqref{d_sol} the solution of~\eqref{Msquare} yields
\begin{equation}
     \matr{X}=\matr{W}^T\matr{X}^\prime\matr{W}. \label{sol_sqrt}
\end{equation}
Using~\eqref{sol_sqrt} and~\eqref{Msquare_d} to compute $\matr{X}^2$, it is straightforward to prove that~\eqref{sol_sqrt} is a solution of~\eqref{Msquare}
\begin{multline}
    \matr{X}^2=\matr{W}^T\matr{X}^\prime\matr{W}\matr{W}^T\matr{X}^\prime\matr{W} =\\
    =\matr{W}^T\matr{X}^\prime\matr{X}^\prime\matr{W}=\matr{W}^T\matr{B}_d\matr{W}=\matr{B}. \label{proof}
\end{multline}

\newpage
\bibliographystyle{unsrt}
\bibliography{Refs.bib}

@misc{lingua_data_2026,
	title = {Data repository for the paper: "{Solving} the inverse parametric problem", {Zenodo}, \url{https://doi.org/10.5281/zenodo.18938314}},
	shorttitle = {Data repository for the paper},
	url = {https://zenodo.org/records/18938314},
	doi = {10.5281/zenodo.18938314},
	language = {eng},
	urldate = {2026-03-11},
	publisher = {Zenodo},
	author = {Lingua, Fabio and Cortinovis, Michele and Haviland, David B.},
	month = mar,
	year = {2026},
}

@article{yokoyama_ultra-large-scale_2013,
	title = {Ultra-large-scale continuous-variable cluster states multiplexed in the time domain},
	volume = {7},
	copyright = {2013 Springer Nature Limited},
	issn = {1749-4893},
	url = {https://www.nature.com/articles/nphoton.2013.287},
	doi = {10.1038/nphoton.2013.287},
	language = {en},
	number = {12},
	urldate = {2023-11-10},
	journal = {Nature Photonics},
	publisher = {Nature Publishing Group},
	author = {Yokoyama, Shota and Ukai, Ryuji and Armstrong, Seiji C. and Sornphiphatphong, Chanond and Kaji, Toshiyuki and Suzuki, Shigenari and Yoshikawa, Jun-ichi and Yonezawa, Hidehiro and Menicucci, Nicolas C. and Furusawa, Akira},
	month = dec,
	year = {2013},
	note = {Number: 12},
	pages = {982--986},
}

@article{caves_quantum_2012,
	title = {Quantum limits on phase-preserving linear amplifiers},
	volume = {86},
	url = {https://link.aps.org/doi/10.1103/PhysRevA.86.063802},
	doi = {10.1103/PhysRevA.86.063802},
	number = {6},
	urldate = {2024-11-30},
	journal = {Physical Review A},
	publisher = {American Physical Society},
	author = {Caves, Carlton M. and Combes, Joshua and Jiang, Zhang and Pandey, Shashank},
	month = dec,
	year = {2012},
	pages = {063802},
}

@article{devaney_nonuniqueness_1978,
	title = {Nonuniqueness in the inverse scattering problem},
	volume = {19},
	issn = {0022-2488},
	url = {https://doi.org/10.1063/1.523860},
	doi = {10.1063/1.523860},
	number = {7},
	urldate = {2026-03-09},
	journal = {Journal of Mathematical Physics},
	author = {Devaney, A. J.},
	month = jul,
	year = {1978},
	pages = {1526--1531},
}

@inproceedings{defacio_nonuniqueness_1992,
	title = {Nonuniqueness in direct and inverse electromagnetic scattering theory},
	volume = {1767},
	url = {https://www.spiedigitallibrary.org/conference-proceedings-of-spie/1767/0000/Nonuniqueness-in-direct-and-inverse-electromagnetic-scattering-theory/10.1117/12.139035.full},
	doi = {10.1117/12.139035},
	urldate = {2026-03-09},
	booktitle = {Inverse {Problems} in {Scattering} and {Imaging}},
	publisher = {SPIE},
	author = {DeFacio, Brian and Kim, S. H.},
	month = dec,
	year = {1992},
	pages = {21--30},
}

@article{del_hougne_virtual_2025,
	title = {Virtual {VNA}: {Minimal}-{Ambiguity} {Scattering} {Matrix} {Estimation} {With} a {Fixed} {Set} of “{Virtual}” {Load}-{Tunable} {Ports}},
	volume = {74},
	issn = {1557-9662},
	shorttitle = {Virtual {VNA}},
	url = {https://ieeexplore.ieee.org/document/10929660},
	doi = {10.1109/TIM.2025.3551897},
	urldate = {2026-03-09},
	journal = {IEEE Transactions on Instrumentation and Measurement},
	author = {del Hougne, Philipp},
	year = {2025},
	pages = {1--19},
}

@article{dersy_reconstructing_2024,
	title = {Reconstructing {S}-matrix {Phases} with {Machine} {Learning}},
	volume = {2024},
	issn = {1029-8479},
	url = {https://doi.org/10.1007/JHEP05(2024)200},
	doi = {10.1007/JHEP05(2024)200},
	language = {en},
	number = {5},
	urldate = {2026-03-06},
	journal = {Journal of High Energy Physics},
	author = {Dersy, Aurélien and Schwartz, Matthew D. and Zhiboedov, Alexander},
	month = may,
	year = {2024},
	pages = {200},
}

@article{kronig_supplementary_1946,
	title = {A supplementary condition in {Heisenberg}'s theory of elementary particles},
	volume = {12},
	issn = {0031-8914},
	url = {https://www.sciencedirect.com/science/article/pii/S0031891446800788},
	doi = {10.1016/S0031-8914(46)80078-8},
	number = {8},
	urldate = {2026-03-06},
	journal = {Physica},
	author = {Kronig, R.},
	month = nov,
	year = {1946},
	pages = {543--544},
}

@article{schutzer_connection_1951,
	title = {On the {Connection} of the {Scattering} and {Derivative} {Matrices} with {Causality}},
	volume = {83},
	url = {https://link.aps.org/doi/10.1103/PhysRev.83.249},
	doi = {10.1103/PhysRev.83.249},
	number = {2},
	urldate = {2026-03-06},
	journal = {Physical Review},
	publisher = {American Physical Society},
	author = {Schützer, Walter and Tiomno, J.},
	month = jul,
	year = {1951},
	pages = {249--251},
}

@article{kampen_causalite_1961,
	title = {Causalité et relations de {Kramers}-{Kronig}},
	volume = {22},
	issn = {0368-3842, 2777-3442},
	url = {http://dx.doi.org/10.1051/jphysrad:01961002203017900},
	doi = {10.1051/jphysrad:01961002203017900},
	language = {fr},
	number = {3},
	urldate = {2026-03-06},
	journal = {Journal de Physique et le Radium},
	publisher = {Revue Générale de l'Electricité},
	author = {Kampen, N. G. Van and Lurçat, François},
	month = mar,
	year = {1961},
	pages = {179--191},
}

@article{menicucci_graphical_2011,
	title = {Graphical calculus for {Gaussian} pure states},
	volume = {83},
	url = {https://link.aps.org/doi/10.1103/PhysRevA.83.042335},
	doi = {10.1103/PhysRevA.83.042335},
	number = {4},
	urldate = {2023-11-27},
	journal = {Physical Review A},
	publisher = {American Physical Society},
	author = {Menicucci, Nicolas C. and Flammia, Steven T. and van Loock, Peter},
	month = apr,
	year = {2011},
	pages = {042335},
}

@article{simon_gaussian_1988,
	title = {Gaussian pure states in quantum mechanics and the symplectic group},
	volume = {37},
	url = {https://link.aps.org/doi/10.1103/PhysRevA.37.3028},
	doi = {10.1103/PhysRevA.37.3028},
	number = {8},
	urldate = {2025-11-17},
	journal = {Physical Review A},
	publisher = {American Physical Society},
	author = {Simon, R. and Sudarshan, E. C. G. and Mukunda, N.},
	month = apr,
	year = {1988},
	pages = {3028--3038},
}

@article{weedbrook_gaussian_2012,
	title = {Gaussian quantum information},
	volume = {84},
	url = {https://link.aps.org/doi/10.1103/RevModPhys.84.621},
	doi = {10.1103/RevModPhys.84.621},
	number = {2},
	urldate = {2023-11-15},
	journal = {Reviews of Modern Physics},
	publisher = {American Physical Society},
	author = {Weedbrook, Christian and Pirandola, Stefano and García-Patrón, Raúl and Cerf, Nicolas J. and Ralph, Timothy C. and Shapiro, Jeffrey H. and Lloyd, Seth},
	month = may,
	year = {2012},
	pages = {621--669},
}

@article{gardiner_input_1985,
	title = {Input and output in damped quantum systems: {Quantum} stochastic differential equations and the master equation},
	volume = {31},
	shorttitle = {Input and output in damped quantum systems},
	url = {https://link.aps.org/doi/10.1103/PhysRevA.31.3761},
	doi = {10.1103/PhysRevA.31.3761},
	number = {6},
	urldate = {2024-02-05},
	journal = {Physical Review A},
	publisher = {American Physical Society},
	author = {Gardiner, C. W. and Collett, M. J.},
	month = jun,
	year = {1985},
	pages = {3761--3774},
}

@article{mollow_quantum_1967,
	title = {Quantum {Theory} of {Parametric} {Amplification}. {I}},
	volume = {160},
	doi = {10.1103/PhysRev.160.1076},
	number = {5},
	journal = {Physical Review},
	author = {Mollow, B. R.},
	year = {1967},
	pages = {1076--1096},
}

@article{bertet_parametric_2006,
	title = {Parametric coupling for superconducting qubits},
	volume = {73},
	doi = {10.1103/PhysRevB.73.064512},
	number = {6},
	journal = {Physical Review B},
	author = {Bertet, P.},
	year = {2006},
}

@article{andersson_squeezing_2022,
	title = {Squeezing and {Multimode} {Entanglement} of {Surface} {Acoustic} {Wave} {Phonons}},
	volume = {3},
	doi = {10.1103/PRXQuantum.3.010312},
	number = {1},
	journal = {PRX Quantum},
	author = {Andersson, Gustav},
	year = {2022},
}

@article{eichler_observation_2011,
	title = {Observation of {Two}-{Mode} {Squeezing} in the {Microwave} {Frequency} {Domain}},
	volume = {107},
	doi = {10.1103/PhysRevLett.107.113601},
	number = {11},
	journal = {Physical Review Letters},
	author = {Eichler, C.},
	year = {2011},
}

@article{mallet_quantum_2011,
	title = {Quantum {State} {Tomography} of an {Itinerant} {Squeezed} {Microwave} {Field}},
	volume = {106},
	doi = {10.1103/PhysRevLett.106.220502},
	number = {22},
	journal = {Physical Review Letters},
	author = {Mallet, F.},
	year = {2011},
}

@article{chen_experimental_2014,
	title = {Experimental {Realization} of {Multipartite} {Entanglement} of 60 {Modes} of a {Quantum} {Optical} {Frequency} {Comb}},
	volume = {112},
	copyright = {http://link.aps.org/licenses/aps-default-license},
	issn = {0031-9007, 1079-7114},
	url = {https://link.aps.org/doi/10.1103/PhysRevLett.112.120505},
	doi = {10.1103/PhysRevLett.112.120505},
	language = {en},
	number = {12},
	urldate = {2025-09-29},
	journal = {Physical Review Letters},
	author = {Chen, Moran and Menicucci, Nicolas C. and Pfister, Olivier},
	month = mar,
	year = {2014},
	pages = {120505},
}

@article{pfister_multipartite_2004,
	title = {Multipartite continuous-variable entanglement from concurrent nonlinearities},
	volume = {70},
	copyright = {http://link.aps.org/licenses/aps-default-license},
	issn = {1050-2947, 1094-1622},
	url = {https://link.aps.org/doi/10.1103/PhysRevA.70.020302},
	doi = {10.1103/PhysRevA.70.020302},
	language = {en},
	number = {2},
	urldate = {2025-09-29},
	journal = {Physical Review A},
	author = {Pfister, Olivier and Feng, Sheng and Jennings, Gregory and Pooser, Raphael and Xie, Daruo},
	month = aug,
	year = {2004},
	pages = {020302},
}

@article{yurke_squeezed-state_1987,
	title = {Squeezed-state generation using a {Josephson} parametric amplifier},
	copyright = {© 1987 Optical Society of America},
	url = {https://opg.optica.org/josab/abstract.cfm?uri=josab-4-10-1551},
	doi = {10.1364/JOSAB.4.001551},
	language = {EN},
	urldate = {2025-09-29},
	journal = {JOSA B, Vol. 4, Issue 10, pp. 1551-1557},
	publisher = {Optica Publishing Group},
	author = {Yurke, B.},
	month = oct,
	year = {1987},
}

@article{cerullo_ultrafast_2003,
	title = {Ultrafast optical parametric amplifiers},
	volume = {74},
	issn = {0034-6748, 1089-7623},
	url = {https://pubs.aip.org/rsi/article/74/1/1/110869/Ultrafast-optical-parametric-amplifiers},
	doi = {10.1063/1.1523642},
	language = {en},
	number = {1},
	urldate = {2025-09-29},
	journal = {Review of Scientific Instruments},
	author = {Cerullo, Giulio and De Silvestri, Sandro},
	month = jan,
	year = {2003},
	pages = {1--18},
}

@article{liao_parametric_2011,
	title = {Parametric generation of quadrature squeezing of mirrors in cavity optomechanics},
	volume = {83},
	copyright = {http://link.aps.org/licenses/aps-default-license},
	issn = {1050-2947, 1094-1622},
	url = {https://link.aps.org/doi/10.1103/PhysRevA.83.033820},
	doi = {10.1103/PhysRevA.83.033820},
	language = {en},
	number = {3},
	urldate = {2025-09-29},
	journal = {Physical Review A},
	author = {Liao, Jie-Qiao and Law, C. K.},
	month = mar,
	year = {2011},
	pages = {033820},
}

@article{petrovnin_generation_2023,
	title = {Generation and {Structuring} of {Multipartite} {Entanglement} in a {Josephson} {Parametric} {System}},
	volume = {6},
	copyright = {© 2022 The Authors. Advanced Quantum Technologies published by Wiley-VCH GmbH},
	issn = {2511-9044},
	url = {https://onlinelibrary.wiley.com/doi/abs/10.1002/qute.202200031},
	doi = {10.1002/qute.202200031},
	language = {en},
	number = {1},
	urldate = {2023-11-10},
	journal = {Advanced Quantum Technologies},
	author = {Petrovnin, Kirill Viktorovich and Perelshtein, Michael Romanovich and Korkalainen, Tero and Vesterinen, Visa and Lilja, Ilari and Paraoanu, Gheorghe Sorin and Hakonen, Pertti Juhani},
	year = {2023},
	pages = {2200031},
}

@article{jolin_multipartite_2023,
	title = {Multipartite {Entanglement} in a {Microwave} {Frequency} {Comb}},
	volume = {130},
	issn = {0031-9007, 1079-7114},
	url = {https://link.aps.org/doi/10.1103/PhysRevLett.130.120601},
	doi = {10.1103/PhysRevLett.130.120601},
	language = {en},
	number = {12},
	urldate = {2023-10-09},
	journal = {Physical Review Letters},
	author = {Jolin, Shan W. and Andersson, Gustav and Hernández, J. C. Rivera and Strandberg, Ingrid and Quijandría, Fernando and Aumentado, José and Borgani, Riccardo and Tholén, Mats O. and Haviland, David B.},
	month = mar,
	year = {2023},
	pages = {120601},
}

@article{esposito_observation_2022,
	title = {Observation of {Two}-{Mode} {Squeezing} in a {Traveling} {Wave} {Parametric} {Amplifier}},
	volume = {128},
	issn = {0031-9007, 1079-7114},
	url = {https://link.aps.org/doi/10.1103/PhysRevLett.128.153603},
	doi = {10.1103/PhysRevLett.128.153603},
	language = {en},
	number = {15},
	urldate = {2025-09-25},
	journal = {Physical Review Letters},
	author = {Esposito, Martina and Ranadive, Arpit and Planat, Luca and Leger, Sébastien and Fraudet, Dorian and Jouanny, Vincent and Buisson, Olivier and Guichard, Wiebke and Naud, Cécile and Aumentado, José and Lecocq, Florent and Roch, Nicolas},
	month = apr,
	year = {2022},
	pages = {153603},
}

@article{lecocq_nonreciprocal_2017,
	title = {Nonreciprocal {Microwave} {Signal} {Processing} with a {Field}-{Programmable} {Josephson} {Amplifier}},
	volume = {7},
	copyright = {http://link.aps.org/licenses/aps-default-license},
	issn = {2331-7019},
	url = {https://link.aps.org/doi/10.1103/PhysRevApplied.7.024028},
	doi = {10.1103/PhysRevApplied.7.024028},
	language = {en},
	number = {2},
	urldate = {2025-09-25},
	journal = {Physical Review Applied},
	author = {Lecocq, F. and Ranzani, L. and Peterson, G. A. and Cicak, K. and Simmonds, R. W. and Teufel, J. D. and Aumentado, J.},
	month = feb,
	year = {2017},
	pages = {024028},
}

@article{castellanos-beltran_amplification_2008,
	title = {Amplification and squeezing of quantum noise with a tunable {Josephson} metamaterial},
	volume = {4},
	issn = {1745-2473, 1745-2481},
	url = {https://www.nature.com/articles/nphys1090},
	doi = {10.1038/nphys1090},
	language = {en},
	number = {12},
	urldate = {2025-09-25},
	journal = {Nature Physics},
	author = {Castellanos-Beltran, M. A. and Irwin, K. D. and Hilton, G. C. and Vale, L. R. and Lehnert, K. W.},
	month = dec,
	year = {2008},
	pages = {929--931},
}

@article{malnou_optimal_2018,
	title = {Optimal {Operation} of a {Josephson} {Parametric} {Amplifier} for {Vacuum} {Squeezing}},
	volume = {9},
	issn = {2331-7019},
	url = {https://link.aps.org/doi/10.1103/PhysRevApplied.9.044023},
	doi = {10.1103/PhysRevApplied.9.044023},
	language = {en},
	number = {4},
	urldate = {2025-09-25},
	journal = {Physical Review Applied},
	author = {Malnou, M. and Palken, D. A. and Vale, Leila R. and Hilton, Gene C. and Lehnert, K. W.},
	month = apr,
	year = {2018},
	pages = {044023},
}

@article{castellanos-beltran_widely_2007,
	title = {Widely tunable parametric amplifier based on a superconducting quantum interference device array resonator},
	volume = {91},
	issn = {0003-6951, 1077-3118},
	url = {https://pubs.aip.org/apl/article/91/8/083509/326765/Widely-tunable-parametric-amplifier-based-on-a},
	doi = {10.1063/1.2773988},
	language = {en},
	number = {8},
	urldate = {2025-09-25},
	journal = {Applied Physics Letters},
	author = {Castellanos-Beltran, M. A. and Lehnert, K. W.},
	month = aug,
	year = {2007},
	pages = {083509},
}

@misc{namdar_spectro-temporally_2025,
	title = {Spectro-temporally tailored {Non}-{Gaussian} {Quantum} {Operations} in {Thin}-{Film} {Waveguides}},
	url = {http://arxiv.org/abs/2508.04578},
	doi = {10.48550/arXiv.2508.04578},
	urldate = {2025-09-25},
	publisher = {arXiv},
	author = {Namdar, Peter and Folge, Patrick and Lopetegui, Carlos E. and Babel, Silia and Brecht, Benjamin and Silberhorn, Christine and Parigi, Valentina},
	month = aug,
	year = {2025},
	note = {arXiv:2508.04578 [quant-ph]},
}

@article{aumentado_superconducting_2020,
	title = {Superconducting {Parametric} {Amplifiers}: {The} {State} of the {Art} in {Josephson} {Parametric} {Amplifiers}},
	volume = {21},
	copyright = {https://ieeexplore.ieee.org/Xplorehelp/downloads/license-information/USG.html},
	issn = {1527-3342, 1557-9581},
	shorttitle = {Superconducting {Parametric} {Amplifiers}},
	url = {https://ieeexplore.ieee.org/document/9134828/},
	doi = {10.1109/MMM.2020.2993476},
	language = {en},
	number = {8},
	urldate = {2024-11-05},
	journal = {IEEE Microwave Magazine},
	author = {Aumentado, Jose},
	month = aug,
	year = {2020},
	pages = {45--59},
}

@article{yurke_lownoise_1996,
	title = {A low‐noise series‐array {Josephson} junction parametric amplifier},
	volume = {69},
	issn = {0003-6951},
	url = {https://doi.org/10.1063/1.116845},
	doi = {10.1063/1.116845},
	number = {20},
	urldate = {2024-11-26},
	journal = {Applied Physics Letters},
	author = {Yurke, B. and Roukes, M. L. and Movshovich, R. and Pargellis, A. N.},
	month = nov,
	year = {1996},
	pages = {3078--3080},
}

@article{naaman_synthesis_2022,
	title = {Synthesis of {Parametrically} {Coupled} {Networks}},
	volume = {3},
	url = {https://link.aps.org/doi/10.1103/PRXQuantum.3.020201},
	doi = {10.1103/PRXQuantum.3.020201},
	number = {2},
	urldate = {2024-11-06},
	journal = {PRX Quantum},
	publisher = {American Physical Society},
	author = {Naaman, Ofer and Aumentado, José},
	month = may,
	year = {2022},
	pages = {020201},
}

@article{ranzani_graph-based_2015,
	title = {Graph-based analysis of nonreciprocity in coupled-mode systems},
	volume = {17},
	issn = {1367-2630},
	url = {https://dx.doi.org/10.1088/1367-2630/17/2/023024},
	doi = {10.1088/1367-2630/17/2/023024},
	language = {en},
	number = {2},
	urldate = {2024-08-13},
	journal = {New Journal of Physics},
	publisher = {IOP Publishing},
	author = {Ranzani, Leonardo and Aumentado, José},
	month = feb,
	year = {2015},
	pages = {023024},
}

@article{lingua_continuous-variable_2025,
	title = {Continuous-{Variable} {Square}-{Ladder} {Cluster} {States} in a {Microwave} {Frequency} {Comb}},
	volume = {134},
	url = {https://link.aps.org/doi/10.1103/PhysRevLett.134.183602},
	doi = {10.1103/PhysRevLett.134.183602},
	number = {18},
	urldate = {2025-07-31},
	journal = {Physical Review Letters},
	publisher = {American Physical Society},
	author = {Lingua, Fabio and Rivera Hernández, J. C. and Cortinovis, Michele and Haviland, David B.},
	month = may,
	year = {2025},
	pages = {183602},
}

@article{hernandez_control_2024,
	title = {Control of multi-modal scattering in a microwave frequency comb},
	volume = {1},
	issn = {2835-0103},
	url = {http://arxiv.org/abs/2402.09068},
	doi = {10.1063/5.0203426},
	language = {en},
	number = {3},
	urldate = {2025-07-31},
	journal = {APL Quantum},
	author = {Hernández, J. C. Rivera and Lingua, Fabio and Jolin, Shan W. and Haviland, David B.},
	month = sep,
	year = {2024},
	note = {arXiv:2402.09068 [quant-ph]},
	pages = {036101},
}

@article{bock_nonreciprocal_2025,
	title = {Nonreciprocal scattering in a microwave frequency comb},
	volume = {24},
	url = {https://link.aps.org/doi/10.1103/kz53-dryz},
	doi = {10.1103/kz53-dryz},
	number = {1},
	urldate = {2025-08-01},
	journal = {Physical Review Applied},
	publisher = {American Physical Society},
	author = {Bock, Christoph L. and Hernández, J.C. Rivera and Lingua, Fabio and Haviland, David B.},
	month = jul,
	year = {2025},
	pages = {014027},
}

@article{arzani_versatile_2018,
	title = {Versatile engineering of multimode squeezed states by optimizing the pump spectral profile in spontaneous parametric down-conversion},
	volume = {97},
	url = {https://link.aps.org/doi/10.1103/PhysRevA.97.033808},
	doi = {10.1103/PhysRevA.97.033808},
	number = {3},
	urldate = {2025-05-06},
	journal = {Physical Review A},
	publisher = {American Physical Society},
	author = {Arzani, Francesco and Fabre, Claude and Treps, Nicolas},
	month = mar,
	year = {2018},
	pages = {033808},
}

@article{landgraf_automated_2025,
	title = {Automated {Discovery} of {Coupled}-{Mode} {Setups}},
	volume = {15},
	url = {https://link.aps.org/doi/10.1103/PhysRevX.15.021038},
	doi = {10.1103/PhysRevX.15.021038},
	number = {2},
	urldate = {2025-05-06},
	journal = {Physical Review X},
	publisher = {American Physical Society},
	author = {Landgraf, Jonas and Peano, Vittorio and Marquardt, Florian},
	month = may,
	year = {2025},
	pages = {021038},
}

\end{document}